\documentclass[runningheads]{llncs}
\usepackage[T1]{fontenc}
\usepackage{hyperref}
\usepackage{graphicx}
\usepackage{doi}
\usepackage{hyperref}
\usepackage{epstopdf}
\usepackage{flushend}
\usepackage{tcolorbox}
\usepackage{orcidlink}
\usepackage{xcolor}
\usepackage{soul}
\usepackage{url}
\usepackage{subcaption}
\usepackage{xcolor}
\usepackage{algorithm}
\usepackage{algorithmic}
\usepackage{amsmath}
\definecolor{harelcommentcolor}{RGB}{50, 150, 200} % A nice forest green

\begin{document}

\title{Survival of the Stealthiest: Evolving Low-Entropy Ransomware via Genetic Algorithms}
\author{Efrat Levenberg\inst{1}\orcidID{0009-0001-6846-9599} \and
Kristina Sviazhina\inst{2}\orcidID{0009-0007-6194-2025} \and
Ayelet Butman\inst{2}\orcidID{0009-0007-5786-1986} \and
Pierre Parrend\inst{3}\orcidID{0000-0002-1680-1182} \and
Harel Berger\inst{4}\orcidID{0000-0001-6035-5127}
\\}

\institute{JCT, Israel 
\email{efratlevenberg@gmail.com} \and
HIT, Israel
\email{krissviazhina@gmail.com,ayeletbutman@gmail.com} \and
EPITA; Université de Strasbourg, France
\email{pierre.parrend@epita.fr} \and
Ariel University, Israel
\email{harelb@ariel.ac.il}
}
\maketitle              % typeset the header of the contribution
\begin{abstract}
Traditional ransomware deployment often relies on massive encryption procedure, triggering immediate detection by modern defense systems. This work introduces a paradigm shift in cryptographic attacks by framing ransomware execution as a Search-Based Software Engineering (SBSE) optimization problem. This approach addresses the persistence gap observed in modern threats, where attacks aim to remain undercover for hours rather than minutes. Using a Genetic Algorithm (GA), we optimize data encryption under a hard constraint on the statistical deviation from baseline system activity. We demonstrate that our evolved attack patterns can evade behavioral monitors under fingerprinting techniques. Our results suggest that search-based methods provide a powerful framework for generating evasive malware, highlighting an emerging challenge for automated software defense.

\end{abstract}
\section{Introduction}
Low-noise ransomware typically employs splitting, intermittent, imitating, or low-entropy encryption to evade detection \cite{Zhao2025NDSS}. Low-entropy encryption is the most sophisticated of these, actively controlling data distribution through partial encryption or masking to avoid the sharp statistical transitions that trigger modern defenses. A prominent example is the Rhysida ransomware\footnote{https://www.cisa.gov/news-events/cybersecurity-advisories/aa23-319a}, which integrates entropy-aware state handling into its workflow to bypass behavioral monitors \cite{kim2024methoddecryptingdatainfected}. Despite the maturity of detection research, the strategic optimization of these attack vectors remains under-explored, leaving a critical gap in anticipating the evolution of malware campaigns.

We address this gap by framing stealthy encryption as a Search-Based Software Engineering (SBSE) problem. We propose a novel approach for progressive file encryption that leverages the generative capabilities of Genetic Algorithms (GA) to evolve file-rewriting strategies. By optimizing a fitness function that balances data disruption against anomaly scores, the GA identifies encryption patterns that remain below the thresholds of common entropy-based detectors. Our evaluation demonstrates that this search-based scheme is both effective and efficient, revealing a potent emerging threat that requires resilient detection.

\noindent\textbf{\textit{Availability.} } The complete replication package is available at~\cite{levenberg_2026_20075234}.

% \noindent
% \textbf{Availability. }%
% Our code and results are available via \href{https://anonymous.4open.science/r/Survival-of-the-Stealthiest-702B/README.md}{anonymous repository}.

\section{Related Work}
\label{soa}

Contemporary defenses deploy deep, feature-based behavioral analysis to counter sophisticated ransomware. Frameworks like Minerva leverage file-level I/O features to achieve near-instantaneous detection with 99.45\% accuracy \cite{minerva2025}, while ML-driven models (e.g., Autonomous Feature Resonance and Random Forests) isolate subtle entropy patterns to separate encrypted data from compressed files \cite{feature_resonance2025}. Maintaining high recall rates above 97\%, these systems present a formidable barrier to traditional, high-volume encryption.

While existing literature heavily explores algorithmic evasion and intermittent tactics, research into sustained adversarial persistence remains limited. For instance, techniques like Format-Preserving Encryption (FPE) can neutralize 97\% of entropy-based alerts \cite{casino2025not}, and AI-driven variants like RansomAI utilize Reinforcement Learning to dynamically adjust encryption parameters against static detectors \cite{mahboubi2025battlefield,Assen2023RansomAI}. However, these approaches rarely address long-term persistence via search-based partial encryption. This work builds upon these foundations, shifting the narrative from rapid execution to sustained, stealthy persistence designed to remain undercover across extended temporal horizons.
\section{Methodology}
\label{methodo}
\begin{center}
\noindent\fbox{%
    \parbox{0.95\linewidth}{%
        \textbf{Research Question (RQ):} \textit{Is a partial ransomware attack stealthier than a regular one-shot ransomware?} 
    }%
}
\end{center}

\subsubsection{System Architecture}
The system consists of a client, a server, and an adversary. During the init phase, the server monitors the activity and constructs a behavioral fingerprint ($\mathcal{F}$), establishing a baseline. Security is maintained through an anomaly detection mechanism where the server periodically calculates the deviation ($\Delta$) between real-time behavior and $\mathcal{F}$. A predefined threshold, $\tau$, dictates the allowable variance before an alert is triggered. 
\subsubsection{Attacker Model}
The adversary’s objective is to execute data modification while ensuring $\Delta < \tau$, treating the detection threshold as a dynamic constraint within the search space. We formulate this as a constrained optimization problem using a GA to evolve a strategy $(C, S)$, where $C$ represents the volume of characters encrypted per operation and $S$ denotes the dormant interval between slots. To maximize evasion and avoid user scrutiny, the attacker employs a stochastic policy that exclusively targets files with no active file descriptors. Operating solely on these files prevents I/O conflicts or visible system errors that could alert a user or application. Finally, the attacker monitors a binary detection signal $D \in \{0, 1\}$ as a terminal feedback mechanism. The GA optimizes attack progression as a single objective subject to the hard constraint of system concealment, evolving parameters within bounded limits $C \in [0, C_{max}]$ and $S \in [0, S_{max}]$ to identify the most effective stealthy encryption sequence.

\subsubsection{Genetic Algorithm}
We utilize a GA to navigate the high-dimensional search space of attack configurations and identify the most effective stealthy encryption sequences. The search process evolves a population of strategy vectors $x = (C, S)$, where $C$ represents the volume of characters encrypted per operation and $S$ denotes the dormant interval. Through iterative cycles of binary tournament selection, crossover, and mutation, the GA optimizes a single-objective fitness function defined as the negated encryption volume, $f(x) = -V$, to maximize attack progression. This optimization is subject to the environmental constraint $g(x) = D$, where a terminal detection signal ($D=1$) renders an individual infeasible and halts its evaluation. We distinguish between \textit{transient detection}, where minor encryption spikes trigger temporary anomaly flags that the GA subsequently corrects via optimization, and \textit{terminal detection} ($D=1$), which triggers an immediate process shutdown. Consequently, transient events are logged but do not halt execution. This dual-layer feedback allows the GA to optimize attack progression within the bounds of the terminal constraint.

\subsubsection{Metric} We define \textit{Encryption Progress ($P$)} as the ratio of successfully processed data blocks to the total size of the target directory ($P = \frac{V_{encrypted}}{V_{total}}$). In this context, 100\% progress signifies that every file has been partially encrypted, rendering the entire dataset non-functional for the user.
\iffalse
Our attack is provided in algorithm~\ref{alg:attack_proc}.
\begin{algorithm}
\caption{Partial Encryption Attack Algorithm (GA-based)}
\label{alg:attack_proc}
\begin{algorithmic}[1]
\STATE \textbf{Initialize:} Population $P_0 \gets \{(C, S)_1, \dots, (C, S)_n\}$
\WHILE{generation $g < G_{max}$}
    \FOR{each individual $i \in P_g$}
        \STATE Deploy attack burst with parameters $(C_i, S_i)$
        \STATE \textbf{Listen:} Intercept system status logs for signal $D$
        \IF{$D = 1$}
            \STATE Mark $i$ as \textit{Infeasible} (Constraint Violation)
            \STATE Terminate burst for individual $i$
        \ELSE
            \STATE Calculate fitness $f_1 = \text{Volume}(i)$ \COMMENT{Single Objective: Throughput}
        \ENDIF
    \ENDFOR
    \STATE $P_{g+1} \gets \text{GA\_Update}(P_g)$ \COMMENT{Selection, Crossover, and Mutation}
\ENDWHILE
\end{algorithmic}
\end{algorithm}
\fi

\section{Experimental Setup}
\label{experiments}

The optimization framework uses the \texttt{PyMoo}~\cite{blank2020pymoo} library in a closed-loop process. The experiments were conducted on a virtualized Ubuntu 24.04.3 LTS (Noble Numbat) environment. The system was provisioned with 4GB of RAM (3.8Gi total) and a matching swap partition. The target dataset is a directory containing 30 .txt files with standard ASCII characters. The encryption we used is a simple substitution cipher (i.e., Caesar's cipher). We compare our approach against a Static Attack baseline,  with constant $C = 1$ characters and $S = 60$ seconds. Our GA attack used the same base configuration as a starting point.

The server employs an Isolation Forest to detect anomalies in an 85-feature fingerprint vector, which includes system metrics (CPU, memory, network) and 60 Linux kernel tracepoints (block I/O, scheduler, and writeback operations). 
During a 120s warmup, the server establishes the fingerprint ($\mathcal{F}$) by collecting anomaly scores of normal system behavior; the detection threshold ($\tau$) is then set at the 1st percentile of these scores. Then, this adaptive threshold is updated every 30s using a 0.4 blending factor. The GA is configured with $N=10$, 50 generations, $P_c = 0.9$, and $P_m = 1.0$. Evaluation bursts are limited to 10s, allowing the optimizer to probe the detection boundary before the server’s threshold recalibrates. 
\section{Results}
\label{results}
\begin{figure*}[htbp]
    \centering
    \includegraphics[width=\linewidth]{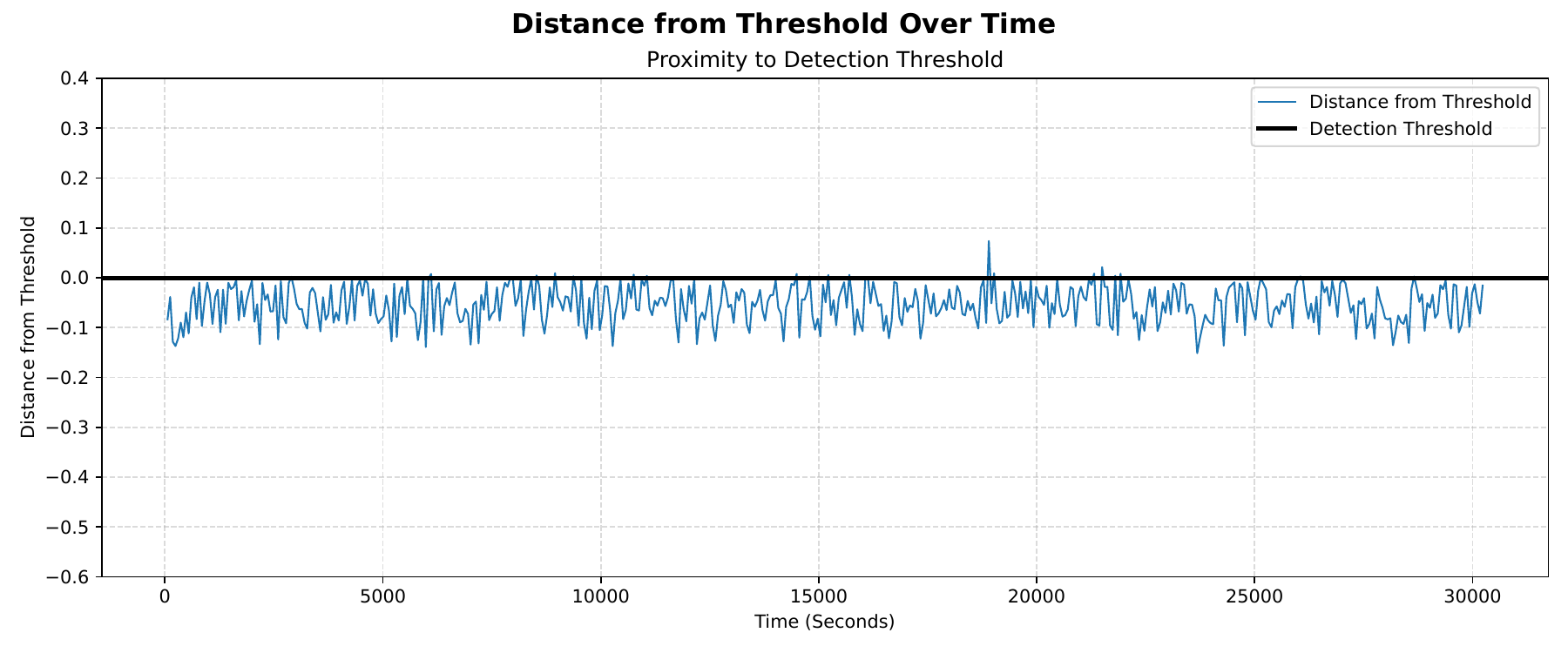}
    \caption{Distance from threshold over time during a 50-generation creeping attack. The attacker actively adapts behavior to remain below the limit.}
    \label{fig:distance_plot}
    \vspace{0.2cm}
    \includegraphics[width=\linewidth]{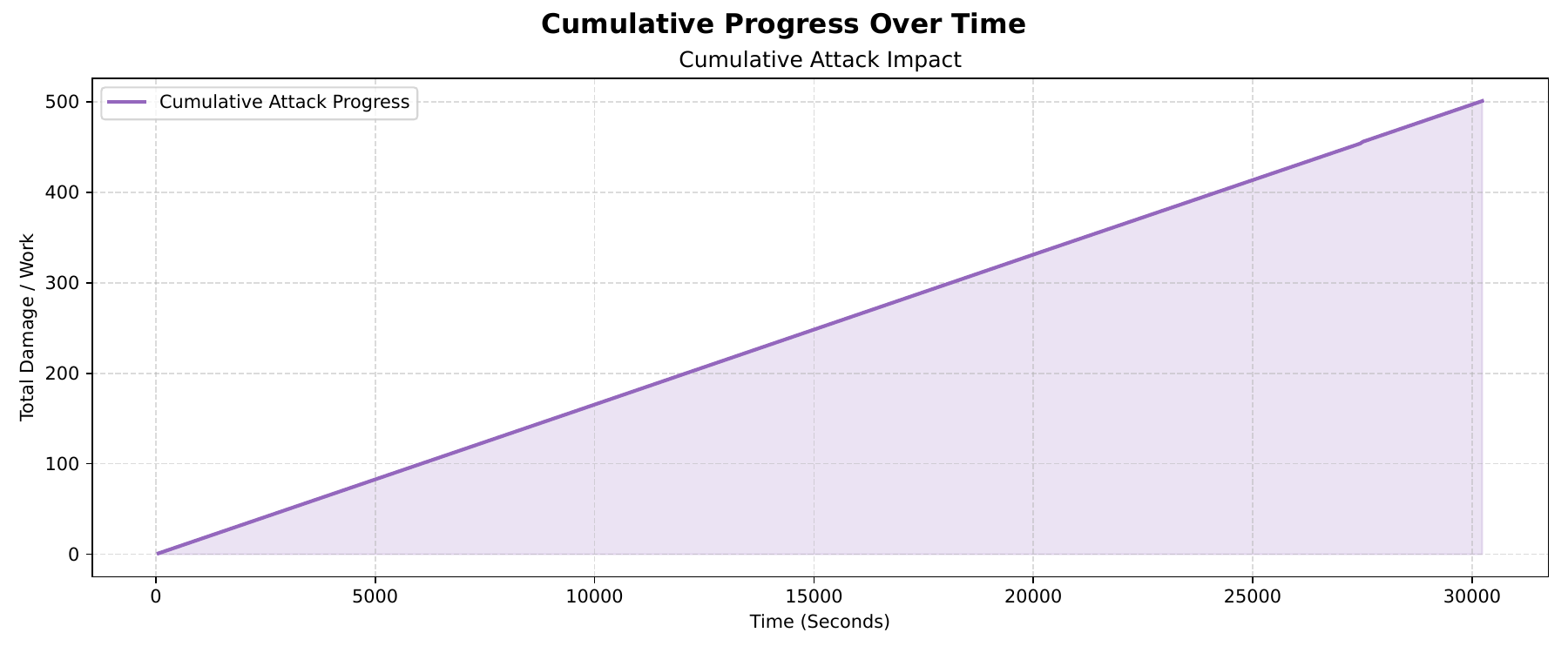}
    \caption{Cumulative attack impact, showing steady damage growth over the 30,000-second window despite the evasive profile.}
    \label{fig:cumulative_plot}
    \vspace{-0.3cm}
\end{figure*}

As illustrated in Figs.~\ref{fig:distance_plot} and \ref{fig:cumulative_plot}, the GA optimizer achieved 100\% dataset coverage over 30,230 seconds. While 21 transient anomaly spikes occurred, they did not trigger a terminal system halt; the GA treated these spikes as fitness penalties, subsequently adapting toward a stealthy throughput of 1 char/60s. Notably, the anomaly score remained aligned with the dynamic threshold, demonstrating the GA’s ability to safely probe the detection boundary without termination.

\begin{table}[htbp]
\centering
\caption{Evaluation: Static Baseline vs. GA-Evolved Attack}
\label{tab:attack_results}
\footnotesize
\begin{tabular}{|l|c|c|}
\hline
\textbf{Metric} & \textbf{Static Attack} & \textbf{GA-Evolved Attack} \\ \hline
Sim. Time & $\sim$3,000s & \textbf{30,230s} \\ \hline
Throughput & 1 char / burst & 1 char / burst \\ \hline
Mean Delay & 60s & \textbf{60.0s (Optimized)} \\ \hline
Detection & 1 (Terminal) & \textbf{21 (Transient)} \\ \hline
Progress (P) & 100\% & \textbf{100\%} \\ \hline
\end{tabular}
\\[3pt]
\scriptsize{*Note: Transient spikes were logged but did not trigger process termination.}
\end{table}

The GA maintained a rhythmic pace that exploited the system's tolerance gap, effectively utilizing the monitoring window without crossing the activation mass. While the static baseline offers a shorter execution window, the GA-evolved approach yields significantly higher reliability in sensitive environments where detection leads to immediate process termination. 

% \noindent\textbf{Summary of Findings (RQ):} Our evaluation demonstrates that a low-and-slow, optimized ransomware profile can remain entirely undetected by advanced anomaly systems for prolonged durations (exceeding 8 hours) while executing complete data modification.
\begin{center}
\noindent\fbox{%
    \parbox{0.95\linewidth}{%
        \textbf{Summary of Findings (RQ):} A partial ransomware can be stealthier than a traditional ransomware, as it remains undetected for a long time ($>$8 hours), through adaptation to the detection systems' behavior.
    }%
}
\end{center}

\section{Threats to Validity}
\label{sec:validity}
While our proof-of-concept demonstrates the potential for evolved stealthy ransomware, several limitations exist. First, the evaluation used a controlled dataset of 30 ASCII files; real-world environments with diverse file types and higher entropy may alter the Isolation Forest's baseline. Second, we assumed a constant system workload, whereas real-world noise from human activity could either mask or expose the attack's signals. Finally, while we simulated an adaptive threshold, further research is required to evaluate persistence against active endpoint defense (EDR) systems that may employ more aggressive defenses.
\section{Conclusion}
\label{concl}
This paper demonstrated that Genetic Algorithms (GAs) can significantly enhance ransomware evasiveness by framing encryption as a constrained optimization problem. Our proof-of-concept evolved a "low-and-slow" profile that sustained an attack for 30,230 seconds ($\sim$8 hours) to achieve 100\% encryption progress without triggering terminal detection. This stability against dynamic, adaptive thresholds highlights a critical vulnerability in current entropy-based defense paradigms. Future work will investigate more sophisticated file-targeting policies, integrate advanced cryptographic schemes like AES-256, and examine how optimized attacks can dynamically adapt to varying system workloads. These findings underscore the urgent need for more resilient, adaptive security frameworks capable of countering search-based adversarial optimization.

\bibliography{biblio}
\bibliographystyle{splncs04}
\end{document}